\documentclass[aps,prb]{revtex4}
\usepackage[T1]{fontenc}
\usepackage[latin1]{inputenc}
\usepackage[american,english]{babel}
\usepackage{graphics}

\makeatletter

\providecommand{\LyX}{L\kern-.1667em\lower.25em\hbox{Y}\kern-.125emX\@}
\newcommand{\lyxline}[1]{
  {#1 \vspace{1ex} \hrule width \columnwidth \vspace{1ex}}
}
\newenvironment{LyXParagraphIndent}[1]%
{
  \begin{list}{}{%
    \setlength\topsep{0pt}%
    \addtolength{\leftmargin}{#1}
    \setlength\parsep{0pt plus 1pt}%
  }
  \item[]
}
{\end{list}}

\usepackage[bbgreekl]{mathbbol}

\makeatother
\begin{document}

\selectlanguage{american}
\title{Homogeneous Nucleation in Inhomogeneous Media \foreignlanguage{english}{I:
Nucleation in a Temperature Gradient}}
\selectlanguage{english}

\author{David Reguera}

\email{davidr@chem.ucla.edu}

\affiliation{Department of Chemistry and Biochemistry, UCLA, 607 Charles E. Young
East Drive, Los Angeles, CA 90095-1569}

\author{J. M. Rub\'{i}}

\thanks{Permanent address: Departament de F\'{i}sica Fonamental, Universitat
de Barcelona, Diagonal 647, 08028 Barcelona, Spain}

\affiliation{Department of Chemistry, Norwegian University of Science and Technology,
7491 Trodheim, Norway}

\begin{abstract}
\selectlanguage{american}
We introduce a general formalism to analyze nucleation phenomena in
inhomogeneous media which considers the influence of the metastable
phase, which is treated as a heat bath in which clusters are embedded,
in the dynamics of the nucleation process. A kinetic equation for
the evolution of the probability density of the clusters, taking into
account possible inhomogeneities in the bath, is derived using mesoscopic
nonequilibrium thermodynamics. From it, we obtain the nucleation rate
and analyze the role played by the metastable phase in the process.
We discuss in particular condensation and polymer crystallization
in a temperature gradient. 
\selectlanguage{english}
\end{abstract}
\maketitle

\section{Introduction}

\selectlanguage{american}
Nucleation is the initial step in many phase transformations, and
thus plays a crucial role in many processes of scientific and technological
importance\cite{oxtoby,chap1:kn:talan,chap1:kn:nuclea,abraham,chap1:kn:Kelton}.
Despite many years of intense investigation, it is still a very challenging
topic due to the inherent difficulties associated with the nonequilibrium
nature of the phenomenon, and to the fact that the most interesting
entity, namely, the embryo of the new phase or \emph{nucleus}, is
a unstable and highly improbable object. Due to these difficulties,
the analysis has been mostly focused on the simplest case of homogeneous,
isothermal and isotropic nucleation. But the real situation under
which nucleation very often occurs does not correspond to those conditions.
One has to take into account that the real process occurs in a medium,
which in general has spatial, thermal or velocity inhomogeneities,
and which in turn may exert a significant influence in the process.

One of the situations in which this fact becomes more evident is in
polymer crystallization. Contrary to the case of solidification of
simple substances, the crystallization of polymers occurs over a wide
range of temperatures. Moreover, in realistic situations, the external
conditions, especially those concerning the temperature, may change
with time and thus alter the process significantly. These characteristics
make it necessary to study the process under nonisothermal and inhomogeneous
conditions. 

Our aim in this paper is to develop a more realistic model of nucleation
and crystallization that considers the possible influence of the medium.
We will focus on the simplest cases in which the medium may affect
the kinetics. These cases involve the presence of spatial inhomogeneities
in the thermodynamic properties or in the conditions under which nucleation
occurs; the presence of temperature gradients; and, finally, the influence
of flow and stress in the nucleation process, a feature that will
be discussed in the accompanying paper (henceforth referred to as
Paper II).

In the case of homogeneous isotropic nucleation, we can leave spatial
dependencies aside since the process occurs identically at any point
within the system. We can thus focus our description on the evolution
of cluster size as a function of external conditions (pressure, temperature,
density). Homogeneity thus justifies the use of a global thermodynamic
approach. 

However, when the system is inhomogeneous, the conditions controlling
condensation and crystallization vary from point to point and therefore
a local description of the process must be considered.

For this purpose, we will divide the whole sample into volume elements
small enough so that, within them, temperature, pressure, density,
etc. may be considered homogeneous, but large enough to guarantee
that thermodynamics can still be applied. In terms of length scales,
this means that the typical length of the volume element must be much
smaller than the gradients present at the system and much larger than
any length at the molecular scale.

It is important to remark, that nucleation involves two clearly different
length scales of respective macroscopic and mesoscopic natures. The
macroscopic scale is the one on which thermodynamic quantities such
as pressure, temperature, density, etc. vary, and these quantities
can be considered as uniform within each elementary cell, even though
they may change along the sample. On the other hand, nucleation occurs
on a mesoscopic scale. 

For that reason, mesoscopic nonequilibrium thermodynamics (MNET) \cite{chap5:temp}-\cite{chap5:ivan}
becomes the natural framework within which analyze the process. MNET
is a powerful, and systematic method to describe the kinetics of nonequilibrium
processes occurring on the mesoscopic scale. In its present formulation
\cite{chap2:entropic}, it constitutes a generalization of the standard
nonequilibrium thermodynamics, capable of describing the kinetics
of mesoscopic entities in terms of arbitrary variables of state or
degrees of freedom. It has been applied successfully to many out-of-equilibrium
processes, and also to the kinetics of homogeneous nucleation \cite{chap5:16,chap2:kramers-like,chap4:noneq-trans-rot}. 

The key ingredient in extending the formalism to the present situation
(i.e. an inhomogeneous medium) is to carry out a local description
resorting to a reference state in which local equilibrium\cite{chap5:kn:degroot}
is assumed to hold. Starting from this reference state we use the
MNET formalism in order to obtain the Fokker-Planck equation describing
the dynamics of the nucleation process in this inhomogeneous medium.
The resulting Fokker-Planck equation provides a complete description
of the system at the mesoscopic level. However, it contains more information
than is accessible from experiments, performed on a macroscopic scale.
We then describe the macroscopic evolution of the system using hydrodynamic
equations, constructed by averaging the relevant quantities with the
probability density occurring in the Fokker-Planck equation. 

In this way, one obtains the kinetic equations governing both nucleation
and transport in the system. The formalism is able to recover some
results of the kinetic theory, but in a more simple and systematic
way, within a nonequilibrium thermodynamics framework. In particular,
we obtain relaxation equations that can describe the short-time inertial
regime of the dynamics. However, for nucleation, this short-time regime
is usually not relevant, hence we focus our analysis on a long-time
regime where diffusion is the dominant mechanism of transport.

For simplicity, we will not consider angular variables. These can
be easily incorporated into our scheme, but their inclusion complicates
the description of the process. For the sake of concreteness, we investigate
in detail two interesting cases. In the first situation, analyzed
in the present paper, we derive the kinetic equations governing nucleation
in spatially inhomogeneous systems under the presence of a temperature
gradient. The second situation, discussed separately in Paper II,
involves nucleation in a stationary flow, representative of the influence
of stresses or mechanical processing. These examples resemble situations
of practical interest. Using the results obtained within our theoretical
framework, we finally discuss their implications for both condensation
experiments and polymer crystallization \cite{chap5:libropoly}.

\section{Nucleation under Spatially Inhomogeneous Non-Isothermal Conditions
\label{Grad T}}

Consider a nucleation process taking place in a non-isothermal situation
where the metastable phase, globally at rest, is subjected to conditions
that create a stationary temperature profile \( T({\textbf {{x}}}) \),
where \( {\textbf {{x}}} \) is the position coordinate. 

The presence of spatial inhomogeneities makes it necessary to perform
a local description in terms of \( f(n,{\textbf {{x}}},{\textbf {{u}}},t)/N \),
the probability density of finding a cluster of \( n\in (n,n+dn) \)
molecules at position \( {\textbf {{x}}}\in ({\textbf {{x}}},{\textbf {{x}}}+d{\textbf {{x}}}) \),
with velocity \( {\textbf {{u}}}\in ({\textbf {{u}}},{\textbf {{u}}}+d{\textbf {{u}}}) \),
at time \( t \). \( N \) is the total number of clusters in the
system. The whole system is then characterized thermodynamically by
the local energy density \emph{per cluster} \( e({\textbf {{x}}},t) \),
the entropy density \emph{per cluster} \( s({\textbf {{x}}},t) \),
and the number density \begin{equation}
\label{densidad}
\rho ({\textbf {{x}}},t)=\rho _{m}+\rho _{c}\equiv \rho _{m}+\int f(n,{\textbf {{x}}},{\textbf {{u}}},t)d{\textbf {{u}}}dn,
\end{equation}
 where \( \rho _{m} \) is the density of the heat bath, assumed constant. 

Entropy variations of the system, can be expressed in terms of the
corresponding local thermodynamic quantities through the Gibbs equation
(see Appendix of Paper II)

\begin{equation}
\label{Gibbs_eq}
\delta \left( \rho s\right) =\frac{\delta \left( \rho e\right) }{T}-\frac{1}{T}\int \mu (n,{\textbf {{x}}},{\textbf {{u}}},t)\, \delta f(n,{\textbf {{x}}},{\textbf {{u}}},t)\, d{\textbf {{u}}}dn,
\end{equation}
where \( \mu (n,{\textbf {{x}}},{\textbf {{u}}},t) \) is a generalized
chemical potential. Notice that the last term in the previous equation
is reminiscent of the corresponding one for a mixture in which the
different components would be specified by the pair of indices \( n \)
and \( {\textbf {{u}}} \). The generalized chemical potential can
be specified through the use of the Gibbs' entropy postulate. This
formula establishes the connection between statistical mechanics and
thermodynamics through the definition of the entropy in terms of a
probability distribution function, i.e.

\begin{equation}
\label{local G postulate}
\rho s=-k_{B}\int f\ln \frac{f}{f_{leq}}\, d{\textbf {{u}}}dn+\left( \rho s\right) _{leq},
\end{equation}
 where \( f_{leq} \) and \( \rho s_{leq} \) are, respectively, the
probability distribution and the entropy at local equilibrium (indicated
by the subscript \( leq \)). Variation of this expression yields

\begin{equation}
\label{variation local G postulate}
\delta (\rho s)=-k_{B}\int \delta f\ln \frac{f}{f_{leq}}\, d{\textbf {{u}}}dn+\delta (\rho s)_{leq}\, ,
\end{equation}
where the variation of the local equilibrium entropy satisfies the
fundamental equation of thermodynamics 

\begin{equation}
\label{variations eq local G postulate}
\delta (\rho s)_{leq}=\frac{1}{T}\delta (\rho e)_{leq}-\mu _{leq}\delta \rho _{leq},
\end{equation}
 and \( \mu _{leq}({\textbf {{x}}}) \) denotes the local equilibrium
chemical potential which is independent of \( n \) and \( {\textbf {{u}}} \)
, but in general will depend on position. By comparison of the previous
equation with the Gibbs equation (\ref{Gibbs_eq}), one obtains the
expression for the generalized chemical potential 

\begin{equation}
\label{mu f/feq}
\mu (n,{\textbf {{u}}},{\textbf {{x}}},t)=k_{B}T\ln \frac{f}{f_{leq}}+\mu _{leq}({\textbf {{x}}}),
\end{equation}
essentially the chemical potential for a species identified by the
indices \( n \) and \( {\textbf {{u}}} \).

At local equilibrium, the system, viewed as a gas of clusters in the
metastable phase, is described at each position by an equilibrium
distribution function of the internal variables (i.e velocities and
sizes) in which the temperature is that at \( {\textbf {{x}}} \).
Consequently, the local equilibrium distribution is just the probability
of occurrence of an equilibrium fluctuation represented by a cluster
of \( n \) particles and velocity \( {\textbf {{u}}} \) \cite{chap5:kn:landau}

\begin{equation}
\label{distribucion de equilibrio gradT}
f_{leq}(n,{\textbf {{u}}},{\textbf {{x}}},t)=\exp \left( \frac{\mu _{leq}-C(n,\bf {u})}{k_{B}T}\right) .
\end{equation}

The quantity \( C(n,{\textbf {{u}}}) \) denotes the free energy cost
of formation of a cluster of size \( n \), velocity \( {\textbf {{u}}} \)
and mass \( m(n)=m_{1}n \), and is given by

\begin{equation}
\label{barrera grad T}
C(n,{\textbf {{u}}})=\Delta G(n)+\frac{1}{2}m(n){\textbf {{u}}}^{2}\, ,
\end{equation}
 where \( \Delta G(n) \) represents the free energy of formation
of a cluster of size \( n \) \emph{at rest} and the second term is
its kinetic energy. By using the local equilibrium distribution (\ref{distribucion de equilibrio gradT}),
the chemical potential can be written as  \begin{equation}
\label{mu}
\mu (n,{\textbf {{x}}},{\textbf {{u}}},t)=k_{B}T\ln f(n,{\textbf {{x}}},{\textbf {{u}}},t)+C(n,{\textbf {{u}}})\, ,
\end{equation}
which corresponds to that of an ideal system since interactions between
clusters have been neglected. 

The next step is to formulate the conservation laws governing the
evolution of the relevant quantities of the system, namely the probability
density \( f \), the internal energy \( e({\textbf {{x}}},t) \),
and the entropy \( s({\textbf {{x}}},t) \). In absence of external
forces, the continuity equation for \( f \) may in general be written
as\begin{eqnarray}
\frac{\partial f}{\partial t}=-{\textbf {{u}}}\cdot \nabla f-{\partial \over \partial {\textbf {{u}}}}\cdot {\textbf {{J}}}_{u}-\frac{\partial J_{n}}{\partial n}\, , & \label{continuidad} 
\end{eqnarray}
 where \( J_{n} \) is the flux of clusters in size space, and \( {\textbf {{J}}}_{u} \)
is a new current resulting from the interaction of the clusters with
the heat bath. 

The conservation law for the local energy density is \cite{chap5:kn:degroot,chap5:temp}
\begin{eqnarray}
\frac{\partial \rho e}{\partial t}=-\nabla \cdot {\textbf {{J}}}_{q}\, , & \label{balance_energia} 
\end{eqnarray}
 and states that, in the absence of external forces and viscous heating
effects, the \emph{total} internal energy of the fluid element at
\( {\textbf {{x}}} \) can only be altered by a heat flux \( {\textbf {{J}}}_{q} \),. 

The rate of entropy change per unit volume can be evaluated by differentiating
the Gibbs equation (\ref{Gibbs_eq}) with respect to time. Thus

\begin{equation}
\label{evolucin de rhos}
\frac{\partial \rho s}{\partial t}=\frac{1}{T}\frac{\partial \rho e}{\partial t}-{1\over T}\int \mu {\partial f\over \partial t}\, dnd{\textbf {{u}}}.
\end{equation}
Introducing the energy balance (\ref{balance_energia}) and the continuity
equation (\ref{continuidad}), we obtain\begin{eqnarray}
\frac{\partial \rho s}{\partial t}=-\frac{1}{T}\nabla \cdot {\textbf {{J}}}_{q}\quad \qquad \qquad \qquad \qquad \qquad  &  & \nonumber \\
+\frac{1}{T}\int \left( {\textbf {{u}}}\cdot \nabla f+{\partial \over \partial {\textbf {{u}}}}\cdot {\textbf {{J}}}_{u}+{\partial J_{n}\over \partial n}\right) dnd{\textbf {{u}}}. & \label{variacion de s} 
\end{eqnarray}
 Integrating by parts, assuming that the fluxes vanish at the boundaries,
and introducing Eq. (\ref{mu}) for the chemical potential leads to
the following balance equation for the entropy\begin{equation}
\label{balance entropia}
\frac{\partial \rho s}{\partial t}=-\nabla \cdot {\textbf {{J}}}_{s}+\sigma \, ,
\end{equation}
where the entropy flux \( {\textbf {{J}}}_{s} \) is given by

\begin{equation}
\label{flujo entropia}
{\textbf {{J}}}_{s}={1\over T}\, {\textbf {{J}}}'_{q}-k_{B}\int {\textbf {{u}}}f(\ln f-1)dnd{\textbf {{u}}}-\int f{\textbf {{u}}}\, \frac{\left( \Delta G(n)-\Delta H(n)\right) }{T}dnd\bf {u}\, ,
\end{equation}
 and the entropy production \( \sigma  \), which must be positive
semidefinite according to the second law of thermodynamics, is\begin{equation}
\label{produccion entropia}
\sigma =-{1\over T^{2}}\, {\textbf {{J}}}'_{q}\cdot \nabla T-\frac{1}{T}\int {\textbf {{J}}}_{u}\cdot \frac{\partial \mu }{\partial \textbf {{u}}}dnd{\textbf {{u}}}-\frac{1}{T}\int J_{n}\frac{\partial \mu }{\partial n}\, dnd\bf {u}.
\end{equation}
In Eq. (\ref{flujo entropia}), \( \Delta H(n) \) is the enthalpy
of an \( n- \)cluster and \( {\textbf {{J}}}_{q}' \) is the irreversible
heat flux 

\begin{equation}
\label{flujo de calor modificado}
{\textbf {{J}}}_{q}'={\textbf {{J}}}_{q}-\int f{\textbf {{u}}}\left( \Delta H(n)+\frac{1}{2}m(n){\textbf {{u}}}^{2}\right) \, dnd\bf {u}\, ,
\end{equation}
in which the integral represents the transport of heat due to the
diffusion of clusters. In order to obtain the entropy flux we employed
the identity

\begin{equation}
\label{auxiliar}
\int \nabla \cdot \left( {\textbf {{u}}}f\right) \ln f\, dnd{\textbf {{u}}}=\nabla \cdot \left( \int {\textbf {{u}}}f(\ln f-1)\, dnd\bf {u}\right) \, ,
\end{equation}
and the thermodynamic Gibbs-Helmholtz relation \cite{chap5:kn:degroot}

\begin{equation}
\label{entalpia}
\nabla \left( \frac{\Delta G(n)}{T}\right) =-\frac{\Delta H(n)}{T^{2}}\nabla T\, ,
\end{equation}
valid when the system is in mechanical equilibrium.

The entropy production can be interpreted as a sum of products of
conjugate forces and fluxes\cite{chap5:kn:degroot}. In the present
situation, Eq. (\ref{produccion entropia}) allow us to identify the
thermodynamic forces \begin{equation}
\label{fuerzas thermod gradT}
-{1\over T^{2}}\, \nabla T\, ,-\frac{1}{T}\frac{\partial \mu }{\partial \textbf {{u}}}\, ,-\frac{1}{T}\frac{\partial \mu }{\partial n}\, ,
\end{equation}
 conjugate to the heat flux \( {\textbf {{J}}}_{q} \), velocity flux
\( {\textbf {{J}}}_{u} \), and size flux \( J_{n} \), respectively.
Thus it becomes clear that the origin of the thermodynamic forces
are the presence of gradients or, in accordance with Eq. (\ref{mu f/feq}),
variations of the distribution function with respect to its local
equilibrium value.

Now, according to the tenets of nonequilibrium thermodynamics, we
may assume linear relationships between forces and fluxes. Assuming
isotropy, which implies that quantities of different tensorial rank
are not coupled, and locality in the internal space for which currents
at each point (\( n,{\textbf {{u}}}) \) are determined only by the
local properties, one obtains\begin{equation}
\label{ley fenome calor}
{\textbf {{J}}}_{q}'=-{L_{TT}\over T^{2}}\, \nabla T-\int \frac{1}{T}\, L_{Tu}\cdot \frac{\partial \mu }{\partial \textbf {{u}}}\, dnd{\textbf {{u}}}\, ,
\end{equation}

\begin{equation}
\label{ley fenome u shear}
{\textbf {{J}}}_{u}=-{L_{uT}\over T^{2}}\, \nabla T-\frac{1}{T}\, L_{uu}\, \cdot \frac{\partial \mu }{\partial \textbf {{u}}}\, ,
\end{equation}

\begin{equation}
\label{ley fenome gamma shear}
J_{n}=-\frac{1}{T}L_{nn}\, \frac{\partial \mu }{\partial n}\, ,
\end{equation}
where by postulate the phenomenological coefficients satisfy the Onsager
reciprocal relations, e.g. \begin{eqnarray}
L_{Tu}=-L_{uT}\, . & \label{Onsager} 
\end{eqnarray}

It is useful to rewrite the phenomenological coefficients in a more
convenient way. Defining \( \lambda \equiv \frac{L_{TT}}{T^{2}} \)
as the thermal conductivity, \( D_{n}\equiv \frac{k_{B}L_{nn}}{f} \)
as the diffusion coefficient in \( n \)-space, and \( \beta \equiv \frac{mL_{uu}}{fT} \)
and \( \xi \equiv \frac{L_{uT}}{fT} \) as friction coefficients,
and using the expression for the chemical potential, Eq. (\ref{mu}),
the currents can be written as follows

\begin{equation}
\label{flujo de q final}
{\textbf {{J}}}_{q}'=-\lambda \nabla T+\int \xi m\left( f{\textbf {{u}}}+\frac{k_{B}T}{m}\frac{\partial f}{\partial {\bf {u}}}\right) dnd{\textbf {{u}}},
\end{equation}

\begin{equation}
\label{flujo de v final}
{\textbf {{J}}}_{u}=-\xi \frac{f}{T}\nabla T-\beta \left( f{\textbf {{u}}}+\frac{k_{B}T}{m}\frac{\partial f}{\partial \bf {u}}\right) ,
\end{equation}

\begin{equation}
\label{flujo de g final}
J_{n}=-D_{n}\left( \frac{\partial f}{\partial n}+\frac{1}{k_{B}T}\frac{\partial C}{\partial n}f\right) .
\end{equation}

Introducing these expressions into the continuity equation, Eq. (\ref{continuidad}),
we finally obtain the Fokker-Planck equation

\begin{eqnarray}
\frac{\partial f}{\partial t}=-{\textbf {{u}}}\cdot \nabla f+\frac{\partial }{\partial n}\left[ D_{n}\left( \frac{\partial f}{\partial n}+\frac{1}{k_{B}T}\frac{\partial C}{\partial n}f\right) \right] + &  & \nonumber \\
+{\partial \over \partial {\textbf {{u}}}}\cdot \left[ \beta \left( f{\textbf {{u}}}+\frac{k_{B}T}{m}\frac{\partial f}{\partial {\textbf {{u}}}}\right) \right] +{\partial \over \partial {\textbf {{u}}}}\cdot \left( \frac{\xi }{T}f\: \nabla T\right) , & \label{Fokker Planck } 
\end{eqnarray}
that governs the evolution of the inhomogeneous density distribution
of clusters in a bath having a non-uniform temperature distribution.
The transport coefficients \( \beta  \) and \( \xi  \) still remain
unspecified. Their identification must be carried out through a proper
interpretation of the macroscopic relaxation equations derived from
the Fokker-Planck equation. We discuss this issue in the next section.

\section{Homogeneous Nucleation in Spatially Inhomogeneous Systems: Diffusion
Regime\label{homo in inhomo}}

The Fokker-Planck equation, Eq. (\ref{Fokker Planck }), provides
the complete description of the evolution of the probability distribution
of clusters at mesoscopic level. In particular, it can even describe
inertial regimes at the earlier stages of the nucleation process and
retains information about the velocity of clusters. However, the process
of equilibration in velocity space is usually much faster than those
in size or coordinate space. After times much larger than the characteristic
time for velocity relaxation, the system enters the diffusion and
thermal diffusion regime where the evolution is governed by a simpler
set of equations. In that regime, we can describe the system in terms
of the moments of the distribution function, which are related to
the hydrodynamic fields, namely the reduced probability density, defined
as

\begin{equation}
\label{densidad browniana}
f_{c}(n,{\textbf {{x}}},t)=\int f(n,{\textbf {{x}}},{\textbf {{u}}},t)d{\textbf {{u}}}\: ,
\end{equation}
 the velocity in \( (n,{\textbf {{x}}}) \)-space  

\begin{equation}
\label{densidad de momento}
{\textbf {{v}}}_{c}(n,{\textbf {{x}}},t)=f^{-1}_{c}\int {\textbf {{u}}}fd{\textbf {{u}}}\: ,
\end{equation}
and the second moment  

\begin{equation}
\label{definicion del tensor de presiones}
{\cal P}=\int ({\textbf {{u}}}-{\textbf {{v}}}_{c})({\textbf {{u}}}-{\textbf {{v}}}_{c})fd{\textbf {{u}}},
\end{equation}
which corresponds to the kinetic part of the pressure tensor \cite{chap5:kn:degroot,chap5:chapman,chap5:hirschfelder}. 

The evolution equations for these moments follow by taking time derivative
of Eqs. (\ref{densidad browniana})-(\ref{definicion del tensor de presiones})
and using the Fokker-Planck equation (\ref{Fokker Planck }) \textbf{}\cite{chap5:ivan,chap5:inertial}.
The result is the continuity equation

\begin{equation}
\label{continuidad_B}
\frac{\partial f_{c}}{\partial t}=-\nabla \cdot f_{c}{\textbf {{v}}}_{c}-\frac{\partial }{\partial n}\int J_{n}d{\textbf {{u}}}\: ,
\end{equation}
 the equation for the conservation of momentum

\begin{equation}
\label{balance v gradT}
f_{c}\frac{d{\textbf {{v}}}_{c}}{dt}=-\nabla \cdot {\cal P}-\beta f_{c}{\textbf {{v}}}_{c}-\frac{\xi }{T}f_{c}\nabla T-\int ({\textbf {{u}}}-{\textbf {{v}}}_{c})\frac{\partial }{\partial n}J_{n}d{\textbf {{u}}},
\end{equation}
and the equation for the evolution of the pressure tensor

\begin{eqnarray}
\frac{d}{dt}{\cal P}=-\nabla \cdot {\cal Q}-2({\cal P}\cdot \nabla {\textbf {{v}}}_{c})^{s}-{\cal P}\nabla \cdot {\textbf {{v}}}_{c}-2\beta {\cal P}-\frac{2k_{B}T}{m}f_{c}\beta  &  & \nonumber \\
-\int ({\textbf {{u}}}-{\textbf {{v}}}_{c})({\textbf {{u}}}-{\textbf {{v}}}_{c})\frac{\partial }{\partial n}J_{n}d{\textbf {{u}}}. &  & \label{balance p grad T} 
\end{eqnarray}
 In the previous expressions, \( {\cal Q}=\int ({\textbf {{u}}}-{\textbf {{v}}}_{c})({\textbf {{u}}}-{\textbf {{v}}}_{c})({\textbf {{u}}}-{\textbf {{v}}}_{c})fd{\textbf {{v}}} \)
is related with the kinetic part of the heat flux, a superscript \( s \)
means symmetric part of a tensor, and the total or mobile derivative
is defined as \begin{equation}
\label{derivada total}
\frac{d}{dt}\equiv \frac{\partial }{\partial t}+{\textbf {{v}}}_{c}\cdot \nabla \: .
\end{equation}
 In a similar way, we could derive the evolution equations for the
higher-order moments of the distribution, which constitute a coupled
hierarchy of hydrodynamic equations \cite{chap5:ivan,chap5:inertial,chap5:wilemski}.
The closure of the hierarchy can be performed through time-scale considerations
\cite{chap5:ivan}.

The momentum conservation law (\ref{balance v gradT}) enables us
to identify the forces acting on the cluster, and the transport coefficients
\( \beta  \) and \( \xi  \). The third term on the right of that
equation represents a force acting on the cluster, arising from the
presence of the temperature gradient, and giving rise to the thermophoretic
effect or thermophoresis \cite{chap5:thermopho}-\cite{chap5:goldhirsch}.
Its origin can be naively understood on microscopic basis \cite{chap5:thermophEinstein}.
The cluster immersed in a thermal gradient is being hit by particles
of the host fluid. Since collisions with particles from the {}``hot''
side are stronger than those from the {}``cold'' side, a net force
driving the particle toward the cold side appears. The force per unit
mass on a particle is given by \begin{equation}
\label{thermophoresis}
F_{T}=-\frac{\varsigma }{m}\nabla T,
\end{equation}
where \( \varsigma  \) is called the thermophoretic coefficient.
By comparison with the third term on the right of Eq. (\ref{balance v gradT})
we can establish the following identification

\begin{equation}
\label{thermophoretic coefficient}
\xi =\varsigma T.
\end{equation}

The second term on the right of Eq. (\ref{balance v gradT}) can be
identified with the hydrodynamic force exerted by the fluid on the
cluster, with \( \beta  \) playing the role of the friction constant.
This quantity establishes the characteristic relaxation time scale
\( \beta ^{-1} \) for the velocity. It can be estimated using Stokes'
law, \( \beta \simeq \frac{6\pi \eta a}{m} \), which indicates that
its magnitude is very large for small clusters. Consequently, the
discussion of the behavior of the system may be carried out by expanding
the hierarchy of evolution equations for the moments in powers of
\( \beta ^{-1} \). This time scale also motivates the separation
of the dynamics into two well-differentiated regimes: an inertial
regime for \( t\ll \beta ^{-1} \), characterized by the relaxation
of the variables toward the diffusion regime, which is achieved when
\( t\gg \beta ^{-1}. \) 

A remarkable feature of our theory is the fact that it provides a
simple and complete framework for accurately analyzing the dynamics
of mesoscopic systems, even in inertial regimes at very short times.
The set of equations (\ref{continuidad_B})-(\ref{balance p grad T})
which governs the hydrodynamic behavior of the `gas' of clusters,
are analogous to that obtained in kinetic theory \cite{chap5:chapman},\cite{chap5:hirschfelder}.
Notice that the continuity equation, Eq. (\ref{continuidad}), has
a form similar to the Boltzmann transport equation, with the flux
\( {\textbf {{J}}}_{u} \) playing the role of the collision integral.
Therefore, within the framework of nonequilibrium thermodynamics,
our mesoscopic theory is able to reproduce the results of kinetic
theory in a simpler way.

For instance, by retaining subsequent orders in the \( \beta ^{-1} \)
expansion of the continuity equation, one can successively recover
corrections to the diffusion behavior, analogous to the Chapman-Enskog
or Burnett corrections of the Boltzmann equation\cite{chap5:inertial,chap5:ivan}.

The short-time inertial regime is usually not observed in typical
experiments, focused on nucleation phenomena occurring on longer time
scales. It is then sufficient to center the description in the diffusion
regime and to truncate the hierarchy of moments at the level of the
pressure tensor, thus discarding contributions from the heat flux
and higher order moments. We shall retain only the lowest order correction
in \( \beta ^{-1} \).

First the evolution equations for the pressure tensor, Eq. (\ref{balance p grad T}),
and the velocity, Eq. (\ref{balance v gradT}), will be discussed
within the diffusion regime. The terms appearing in the equation for
the pressure tensor (\ref{balance p grad T}), involve different time
scales. In the diffusion regime, i.e. for \( t\gg \beta ^{-1} \),
time derivatives can be neglected when compared with terms proportional
to \( \beta  \). Notice that, in this sense, the divergence term
\( \nabla \cdot {\textbf {{v}}}_{c} \) is essentially a time derivative
(a fact that follows from mass conservation equation), and accordingly
can be neglected. Moreover, velocity relaxation is usually faster
than relaxation in the size-space. Consequently, we can also neglect
the contribution arising from the current \( J_{n} \). Taking all
these considerations into account, the equation for the pressure tensor
then reduces to 

\begin{equation}
\label{TFD para la u}
{\cal P}=\frac{k_{B}T}{m}f_{c}{\mathbb 1},
\end{equation}
 which is that for an ideal gas \cite{chap5:kn:degroot,chap5:temp},
being \( {\mathbb 1} \) the unit tensor. Introducing the value of
the pressure tensor into Eq. (\ref{balance v gradT}) for the evolution
of the velocity yields

\begin{equation}
\label{balance de momento final}
f_{c}\frac{d{\textbf {{v}}}_{c}}{dt}=-\nabla \cdot \left( D_{0}\beta f_{c}{\mathbb 1}\right) -\beta \frac{\xi \beta ^{-1}}{T}f_{c}\nabla T-\beta f_{c}{\textbf {{v}}}_{c}-\int ({\textbf {{u}}}-{\textbf {{v}}}_{c})\frac{\partial }{\partial n}J_{n}d{\textbf {{u}}},
\end{equation}
where \( D_{0}=\frac{k_{B}T}{m\beta } \) is the \emph{spatial} diffusion
coefficient. In the diffusion regime we can again safely neglect the
time derivative and the term proportional to \( J_{n} \). The resulting
equation can be written as

\begin{equation}
\label{corriente difusiva}
{\textbf {{J}}}_{D}=f_{c}{\textbf {{v}}}_{c}=-\nabla \left( D_{0}f_{c}\right) -\frac{\xi \beta ^{-1}}{T}\nabla T\, f_{c},
\end{equation}
which defines the current in the diffusion regime. Inserting previous
expression into the continuity equation (\ref{continuidad_B}), yields

\begin{equation}
\label{FP eq en el regimen difusivo}
\frac{\partial f_{c}}{\partial t}=\nabla \cdot \left( \nabla \left( D_{0}f_{c}\right) +\xi \beta ^{-1}\nabla \ln T\, f_{c}\right) -\frac{\partial }{\partial n}\int J_{n}d{\textbf {{u}}}\: ,
\end{equation}
 where the velocity integral of the current in \( n \)-space is \begin{equation}
\label{integral de J g}
\widetilde{J}_{n}\equiv \int J_{n}d{\textbf {{u}}}=\int \left( \frac{D_{n}}{k_{B}T}\frac{\partial C(n)}{\partial n}f+D_{n}\frac{\partial f}{\partial n}\right) d{\textbf {{u}}}.
\end{equation}

We can reasonably assume that, \( D_{n} \), the diffusion coefficient
in \( n \)-space is approximately independent of the velocity, but
that in general the drift term \( \frac{\partial C}{\partial n} \)
does depend on \( {\textbf {{u}}} \). In fact, from the energy barrier
formula, Eq. (\ref{barrera grad T}), this drift is

\begin{equation}
\label{drift}
\frac{\partial C}{\partial n}=\frac{1}{2}m_{1}u^{2}+\frac{\partial \Delta G(n)}{\partial n}.
\end{equation}

However, in the diffusion regime the system will have achieved equilibration
in velocity space. Therefore the relation 

\begin{equation}
\label{equiparticion}
\int \frac{1}{2}mu^{2}fd{\textbf {{u}}}=\frac{3}{2}k_{B}Tf_{c}\: ,
\end{equation}
 reminiscent of the energy equipartition, holds here as well as in
equilibrium. The integral of the current in size-space can then be
written as\begin{eqnarray}
\widetilde{J}_{n}=D_{n}\left[ \frac{\partial f_{c}}{\partial n}+\frac{1}{k_{B}T}\left( \frac{\partial \Delta G}{\partial n}+\frac{3}{2}\frac{k_{B}T}{n}\right) f_{c}\right]  &  & \nonumber \\
=D_{n}\left[ \frac{\partial f_{c}}{\partial n}+\frac{1}{k_{B}T}\frac{\partial \widetilde{\Delta G}(n)}{\partial n}f_{c}\right] , &  & \label{as} 
\end{eqnarray}
where \( \widetilde{\Delta G}(n) \) is the height of a modified nucleation
barrier whose value is

\begin{equation}
\label{barrera corregida}
\widetilde{\Delta G}(n)=\Delta G(n)+\frac{3}{2}k_{B}T\ln n,
\end{equation}
which includes in an averaged way the effects of the Brownian diffusion
of clusters. This correction was discussed in more detail in Ref.
\cite{chap4:noneq-trans-rot}, where rotational degrees of freedom
were also taken into account.

The evolution equation for the cluster distribution function in spatially
inhomogeneous systems can then finally be expressed as\begin{equation}
\label{FP grad T en el regimen difusivo}
\frac{\partial f_{c}}{\partial t}=\nabla \cdot \left( \nabla \left( D_{0}f_{c}\right) +\xi \beta ^{-1}\nabla \ln T\, f_{c}\right) +\frac{\partial }{\partial n}\left[ D_{n}\left( \frac{\partial f_{c}}{\partial n}+\frac{1}{k_{B}T}\frac{\partial \widetilde{\Delta G}}{\partial n}f_{c}\right) \right] .
\end{equation}

It is convenient to rewrite the spatial flux of clusters as 

\begin{equation}
\label{flujo espacial}
{\textbf {{J}}}_{D}=-D_{0}\nabla f_{c}-\frac{k_{B}}{m\beta }f_{c}\nabla T-\frac{\xi \beta ^{-1}}{T}f_{c}\nabla T=-D_{0}\nabla f_{c}-D_{th}\frac{\nabla T}{T}f_{c},
\end{equation}
 where it is evident that \( J_{D} \) has two contributions: ordinary
diffusion described by Fick's law and a drift term that can be identified
with thermal diffusion. The quantity 

\begin{equation}
\label{thermal diffusion coeff.}
D_{th}=\left( D_{0}+\frac{\xi }{\beta }\right) =D_{0}\left( 1+\frac{\xi m}{k_{B}T}\right) 
\end{equation}
 is the \emph{thermal diffusion coefficient.} Another interesting
quantity is the \emph{thermal diffusion ratio}

\begin{equation}
\label{thermal diffusion ratio}
k_{T}=\frac{D_{th}}{D},
\end{equation}
which measures the importance of thermal diffusion, also known as
Soret effect. Finally, it is important to emphasize that the relation
found in Ref. \cite{chap5:goldhirsch} between the thermophoresis
and the Soret effect, \begin{equation}
\label{relacion entre thermophoresis y soret}
\varsigma =k_{B}(k_{T}-1),
\end{equation}
 is automatically recovered, a fact that follows in a straightforward
manner from Eqs. (\ref{thermophoretic coefficient}), (\ref{thermal diffusion coeff.})
and (\ref{thermal diffusion ratio}).

\subsection{Temperature Evolution}

The above formalism not only prescribes the evolution of the probability
density, but it can also describe the associated evolution of the
temperature field that originates from the balance of internal energy. 

The local internal energy has two contributions: the internal energy
of the clusters and that of the heat bath. If we neglect thermal expansion,
variations of the total internal energy can be related to temperature
variations through the thermodynamic relation \( \rho \delta e=c_{V}\rho \delta T \),
where \( c_{V} \) is the specific heat of the system, bath plus clusters,
at constant volume. Neglecting viscous heating effects, the total
internal energy may vary due to the presence of the heat flux and
also due to the release of latent heat associated with the phase transition.
The heat released in the formation of a cluster of the new stable
phase of size \( n \) is \( \ell m(n)f_{c} \) , where \( \ell  \)
is the latent heat per unit mass. Consequently, the equation governing
the evolution of the temperature field is 

\begin{equation}
\label{variacion de energia total}
\rho c_{V}\frac{dT}{dt}=-\nabla \cdot {\textbf {{J}}}_{q}+\ell \frac{d\rho _{t}}{dt}
\end{equation}
where \( \rho _{t}({\textbf {{x}}},t)=\int m(n)f_{c}dn \) represents
the total density of the new phase (liquid in the case of condensation
or crystallized material in the case of crystal nucleation).

We have seen that the coupling between thermal and diffusion effects
modifies both the diffusion current and the heat flux. To show this
feature it is more convenient to work with the unmodified heat current
because it is usually the quantity measured in experiments

\begin{equation}
\label{flujo de clor exper}
{\textbf {{J}}}_{q}={\textbf {{J}}}_{q}'+\int {\textbf {{u}}}(\Delta H(n)+\frac{1}{2}mu^{2})fdnd\bf {u}.
\end{equation}

Substituting in this expression the equation for the flux, Eq. (\ref{flujo de q final}),
and using \begin{equation}
\label{resultado}
\int {\textbf {{u}}}\frac{1}{2}mu^{2}fdnd{\textbf {{u}}}=\int k_{B}T{\textbf {{J}}}_{D}dn,
\end{equation}
 which is valid in the diffusion regime, we obtain

\begin{equation}
\label{flujo de calor}
{\textbf {{J}}}_{q}=-\lambda \nabla T+\int \left( \Delta H(n)+k_{B}T+\xi m\right) {\textbf {{J}}}_{D}dn.
\end{equation}
Now, employing Eq. (\ref{flujo espacial}) for the diffusion current
\( {\textbf {{J}}}_{D} \), we finally arrive at

\begin{equation}
\label{flujo de calor final}
{\textbf {{J}}}_{q}=-\widetilde{\lambda }\nabla T-\int D_{th}k_{B}T\frac{\nabla f_{c}}{f_{c}}dn+\int \Delta H(n){\textbf {{J}}}_{D}dn,
\end{equation}
 where 

\begin{equation}
\label{conductividad modificada}
\widetilde{\lambda }=\lambda +\int \frac{D^{2}_{th}k_{B}T}{Df_{c}}dn
\end{equation}
 is the modified heat conductivity.

The equation for the evolution of the temperature can finally be expressed
as

\begin{equation}
\label{campo de temperatura}
c_{V}\rho \frac{dT}{dt}=\nabla \cdot \left( \widetilde{\lambda }\nabla T\right) -\nabla \cdot \int \Delta H(n){\textbf {{J}}}_{D}dn+\nabla \cdot \int D_{th}k_{B}T\frac{\nabla f_{c}}{f_{c}}dn+\ell \frac{d\rho _{t}}{dt}.
\end{equation}

From it, we can identify the basic mechanisms responsible for temperature
variations, namely heat conduction, convection, thermal diffusion,
and the release of latent heat in the crystallization process, corresponding
to the different terms on the right hand side of that equation, respectively.
Notice that, by neglecting the contributions arising from the diffusion
of clusters and the modification of the heat conductivity, one can
recover the usual equation for the evolution of temperature field,
namely

\begin{equation}
\label{ecuacion del calor}
\rho _{t}c_{v}\frac{dT}{dt}=\lambda \nabla ^{2}T+\ell \frac{d\rho _{t}}{dt}.
\end{equation}

Once having developed the general formalism describing nucleation
in inhomogeneous media, our aim will be to analyze the influence that
the presence of these inhomogeneities may play in the nucleation process.
This is the topic of the next section.

\subsection{Influence of Diffusion and Thermal Diffusion in Nucleation Experiments}

\subsubsection{Condensation Experiments in Thermal Diffusion Cloud Chambers}

Many systems in which nucleation occurs are spatially nonuniform.
In fact, some experimental setups impose and take advantage of these
inhomogeneities in order to bring about and measure nucleation rates.
For instance, in nucleation experiments in diffusion chambers a temperature
(thermal diffusion chambers) or a velocity gradient (laminar flow
diffusion chambers) is imposed to generate a very narrow region in
which supersaturation exceeds the critical value so that nucleation
occurs. 

Different mechanisms exist through which the presence of inhomogeneities
may influence the nucleation process. On one hand, gradients change
the heat and mass transport in the system as well as the transport
coefficients. Moreover, thermal and diffusion effects may induce convection,
which strongly alters transport in the chamber. In that case, the
conditions of supersaturation and temperature under which nucleation
occurs may not be accurately described in these experiments. This
is an important problem that has been abundantly treated in the literature
\cite{chap5:heist}-\cite{chap5:kane}, and will not be discussed
here. 

On the other hand, there is a direct effect due to the loss of sub-critical
clusters through diffusion and thermal diffusion. If the region in
which nucleation occurs is very narrow, sub-critical clusters may
escape from that region before having had time to grow beyond the
critical size. This mechanism may prevent nucleation or significantly
reduce its rate. We will focus our analysis precisely on this effect,
partially extending previous works by Becker and Reiss \cite{chap5:becker},
and Shi \emph{et al. \cite{chap5:seinfeld}.} 

In real experiments in thermal diffusion cloud chambers, the imposition
of a temperature gradient generates temperature, pressure and supersaturation
profiles inside the chamber (for a representative illustration see
Fig. 1 in Ref. \cite{chap5:npentanol_thermal}). Measurements are
performed under steady state conditions, thus implying that both these
profiles and the nucleation rate are stationary. A salient feature
of this profiles is that, typically, the region in which nucleation
takes place is quite narrow. The activated nature of nucleation implies
that when supersaturation is below the critical value, nucleation
is for all intents and purposes absent. 

\begin{figure}
\selectlanguage{american}
{\centering \resizebox*{1\columnwidth}{!}{\includegraphics{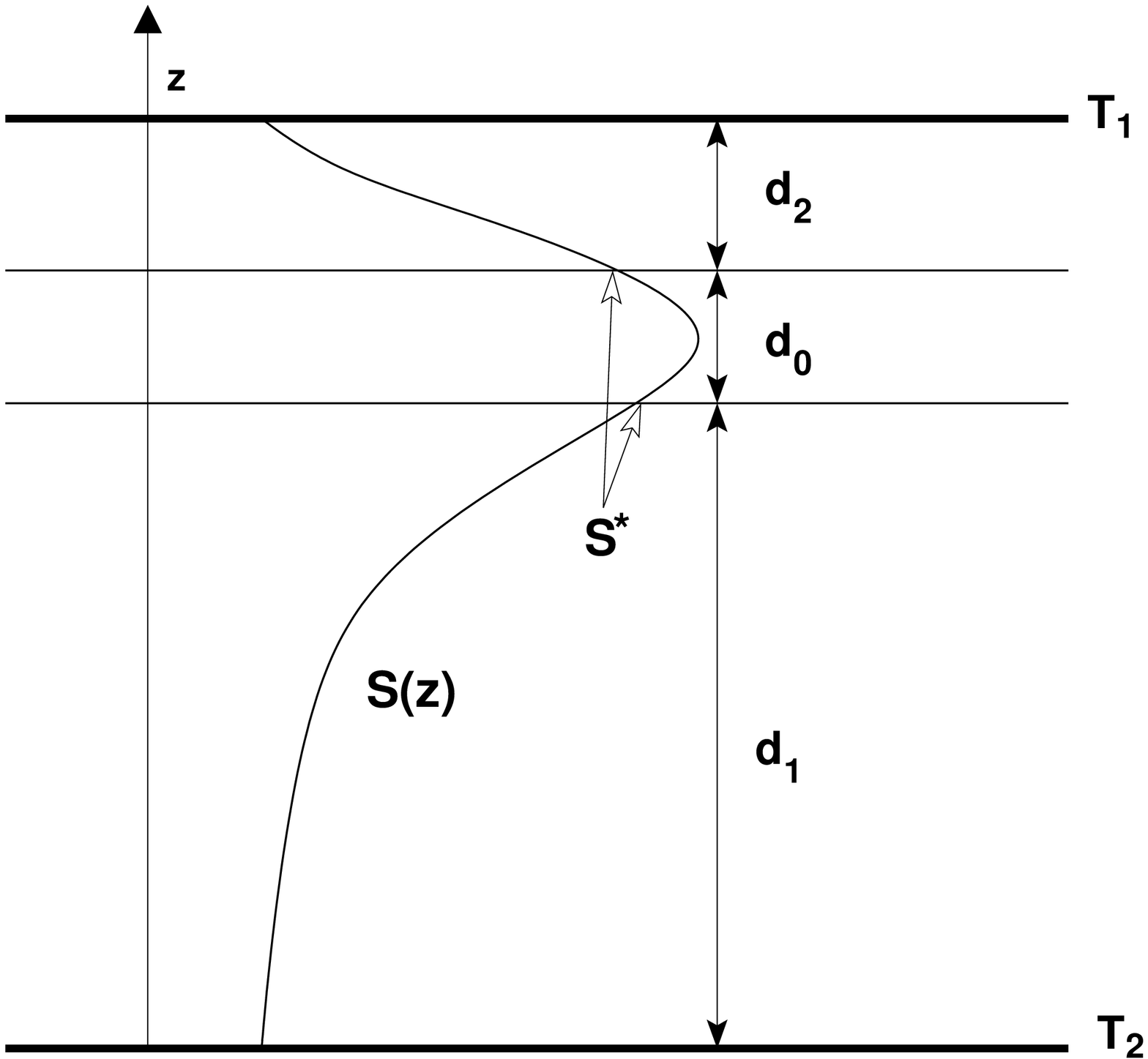}} \par}

\caption{\mbox{}\label{esquema}}
\selectlanguage{english}
\end{figure}

Under these conditions, Eq. (\ref{FP grad T en el regimen difusivo})
can be simplified by assuming that nucleation takes place only in
a small region of thickness \( d_{0} \) (see Fig. \ref{esquema}).

The loss of sub-critical clusters per unit of volume in that region
by diffusion and thermal diffusion can be approximated by 

\begin{equation}
\label{estimacion 1}
loss=\frac{1}{Ad_{0}}\int \nabla \cdot J_{D}d{\textbf {{x}}}\simeq \frac{J_{D}}{d_{0}}=\frac{1}{d_{0}}\left( -D_{0}\nabla f_{c}-D_{th}\frac{\nabla T}{T}f_{c}\right) .
\end{equation}

The saturation ratio decays rapidly on both sides, and due to the
fact that the concentration of clusters depends strongly on supersaturation,
there are virtually no clusters outside that region. If the concentration
of clusters is zero on the boundaries of the chamber, then one can
estimate the gradients for diffusion as \( f_{c}/d_{1} \) and \( f_{c}/d_{2} \)
(see Fig. \ref{esquema} for definitions of \( d_{1} \) and \( d_{2} \))
. Taking these considerations into account the rate of loss of \( n \)-sized
clusters by diffusion and by thermal diffusion is

\begin{equation}
\label{estimacion 2}
-\frac{1}{d_{0}d}D_{0}f_{c}-D_{th}\nabla \ln T\frac{1}{d_{0}}f_{c},
\end{equation}
where \( d^{-1}=d^{-1}_{1}+d^{-1}_{2} \).

It is important to remark that these approximations for the diffusion
and thermal diffusion terms of Eq. (\ref{FP grad T en el regimen difusivo})
are lower bounds that underestimate the actual value, since a sub-critical
cluster will decompose immediately outside the region \( d_{0} \),
and thus the real gradients are consequently stronger. Introducing
these simplifications and the stationary condition into Eq. (\ref{FP grad T en el regimen difusivo}),
we obtain \begin{equation}
\label{Stationary approximate eq}
0=-\frac{\partial }{\partial n}\widetilde{J}_{n}-D_{0}\frac{1}{d\, d_{0}}f_{c}-D_{th}\frac{1}{d_{0}}\frac{d\ln T}{dz}f_{c},
\end{equation}
where \( \widetilde{J}_{n} \) is the flux of clusters in size space. 

Since only order of magnitude estimates of diffusion and thermal diffusion
effects are of interest, we adopt some additional simplifications.
For simplicity, we will not include the nonequilibrium correction
to the nucleation barrier given by Eq. (\ref{barrera corregida}).
This certainly modifies the actual value of the nucleation rate, but
has a weak influence on the diffusion due to its logarithmic dependence
on size. We will then adopt the CNT model for the flux of clusters\cite{chap1:kn:talan},
namely

\begin{equation}
\label{j clas}
\widetilde{J_{n}}=-k^{+}(n)f_{eq}\frac{\partial }{\partial n}\left( \frac{f_{c}}{f_{eq}}\right) ,
\end{equation}
where \( f_{eq}(z,t) \) is the equilibrium distribution of clusters
given by

\begin{equation}
\label{n equilibrio}
f_{eq}(z,t)=N_{1}(t)\exp (n\ln S-\theta n^{2/3}),
\end{equation}
where \( N_{1} \) is the total number of monomers; \( k^{+}(n) \)
is the rate at which an \( n \)-cluster gains monomers, \( \theta =\frac{\sigma s_{1}}{k_{B}T} \)
is the non-dimensional surface tension, being \( \sigma  \) the bulk
surface tension and \( s_{1} \) the area per molecule of the bulk
liquid; and \( S \) is the supersaturation.

It is also convenient to construct a dimensionless version of equation
(\ref{Stationary approximate eq}) by resorting to time-scale considerations.
It is evident that the effect of diffusion will be important when
the time that the cluster takes to diffuse out of the nucleation region
is smaller than the time it takes to grow beyond the critical size.
The spatial and thermal diffusion coefficients, and the rate of growth
of a cluster provide the proper time scales for each of these processes.
The characteristic nucleation relaxation time (see Ref. \cite{chap5:seinfeld})
is \begin{equation}
\label{scala de t nucleation}
\tau _{n}=\frac{\delta ^{2}}{2k^{+}(n^{*})},
\end{equation}
where \( \delta =3(n^{*})^{2/3}\theta ^{-1/2} \) defines the width
of the interval around the critical size \( n^{*}=\left( \frac{2\theta }{3\ln S}\right) ^{3} \)
within which the height of the nucleation barrier has dropped by less
than \( 1k_{B}T \). The diffusion and thermal diffusion time scales
are in turn \( \tau _{B}=\frac{dd_{0}}{D_{0}(n^{*})}, \) and\( \tau _{th}=\frac{d_{0}}{\frac{d\ln T}{dz}D_{th}}, \)
respectively. Therefore, the dimensionless parameters which control
the influence of diffusion and thermal diffusion are \( a=\frac{\tau _{n}}{\tau _{B}}, \)
and \( b=\frac{\tau _{n}}{\tau _{th}}. \) It is then clear, that
a value of \( a \) or \( b \) larger than unity implies that the
time required for a cluster to nucleate is larger than the time it
takes to move out of the nucleating region \( d_{0} \) by diffusion
or thermal diffusion, respectively. In that case a reduction in the
nucleation rate by loss of sub-critical clusters is expected. In contrast,
values of \( a \) and \( b \) smaller than unity indicate that a
cluster in the nucleation region has enough time to become stable
by growing beyond the critical size before it diffuses away. In that
situation, the nucleation rate is not significantly altered by diffusion
or thermal diffusion.  

Introducing these parameters into Eq. (\ref{Stationary approximate eq}),
and defining the reduced distribution function \( y\equiv f_{c}/f_{eq} \),
and size \( x\equiv n/n^{*} \), we obtain the dimensionless expression 

\begin{equation}
\label{estacionaria adimensional }
\varepsilon ^{2}\frac{d^{2}y}{dx^{2}}+\left( \frac{2}{3x}\varepsilon ^{2}+6\left( 1-x^{-1/3}\right) \right) \frac{dy}{dx}-2\left( ax^{-4/3}+bx^{-2/3}\right) y=0,
\end{equation}
where \( \varepsilon \equiv \delta /n^{*} \). This equation is useful
for estimating the effects of the inhomogeneities in the nucleation
rate for different values of the parameters \( a \) and \( b \).

\subsubsection{Numerical Estimates for the Case of Condensation}

To analyze the importance of diffusion and thermal diffusion in real
nucleation experiments, equation (\ref{estacionaria adimensional })
has been solved numerically \cite{chap5:mathematica} for different
values of the parameters \( a \) and \( b \). We have employed the
usual boundary conditions \( y=1 \) for \( n=1 \), and \( y=0 \)
for \( n\rightarrow \infty  \). 

For the case of condensation, the expression for the diffusion and
thermal diffusion coefficients can be borrowed from kinetic theory.
In fact, the Brownian diffusion coefficient of an \( n- \)sized cluster
is approximately given by \cite{chap5:chapman}

\begin{equation}
\label{D teoria cinetica}
D_{0}(n)=\frac{12\pi (k_{B}T)^{3/2}n^{-2/3}}{8P_{tot}\sqrt{2\pi M_{c}}s_{1}},
\end{equation}
where \( P_{tot} \) is the total pressure and \( M_{c} \) is the
molecular mass of the host gas. The thermal diffusion coefficient
can be approximated \cite{chap5:chapman} by 

\begin{equation}
\label{D thermal teoria cinetica}
D_{th}=k_{T}D(n)=\alpha _{T}x_{1}(1-x_{1})D(n),
\end{equation}
where \( \alpha _{T} \) is the thermal diffusion factor, and \( x_{1}=p/P_{tot} \)
is the mole fraction of the nucleating substance. One then obtains

\begin{equation}
\label{a}
a=\frac{27\pi \left( k_{B}T\right) ^{3}}{4p_{eq}S\, P_{tot}s_{1}^{3}\sqrt{\frac{M_{c}}{m_{1}}}}\phi ,
\end{equation}
and

\begin{equation}
\label{b}
b=\frac{27\pi \left( k_{B}T\right) ^{3}}{4p_{eq}S\, P_{tot}s_{1}^{3}\sqrt{\frac{M_{c}}{m_{1}}}}\alpha _{T}x_{1}(1-x_{1})\psi ,
\end{equation}
where \( \phi =(dd_{0})^{-1} \) and \( \psi =\frac{d\ln T}{dz}d^{-1}_{0} \).
In the previous expressions, the parameters that can be changed significantly
are the total pressure, and especially the equilibrium (coexistence)
pressure which decreases exponentially with the inverse temperature.
One can then infer that pressure is the main quantity that determines
the importance of diffusion and thermal diffusion. At low enough pressures,
\( a \) and \( b \) can be of the order of unity , thus indicating
the potential relevance of diffusion and thermal diffusion under rarefied
conditions.

To be more precise, we discuss two particular situations. The first
deals with the influence of diffusion and thermal diffusion in a real
experiment \cite{chap5:npentanol_thermal} in a thermal diffusion
cloud chamber, where n-pentanol is the nucleating substance. The second
deals with the same problem using a model compound that imitates the
properties of a typical sulfinic acid, a substance of great interest
in atmospheric processes. Thermophysical properties of both substances
are listed in Table \ref{propiedades npentanol}. Typical values of
the geometric parameters in a diffusion cloud chamber are \cite{chap5:hung}:
\( d_{0}=0.1h \), \( d_{1}=0.7h \), and \( d_{2}=0.2h \), where
\( h=42.3 \) mm is the height of the chamber, which results in \( \phi =3.6 \),
and \( \psi \sim 0.1 \) for a representative value of \( 50K \)
for the temperature difference between the upper and lower plates. 

Figures \ref{pentanol a} and \ref{pentanol b} present the ratio
between the rate of nucleation in the presence of diffusion and thermal
diffusion effects and in their absence, as a function of the dimensionless
parameters \( a \) and \( b \). Plots have been obtained from numerical
resolution of Eq. (\ref{estacionaria adimensional }) using Mathematica,
and the properties of n-pentanol at \( T=260K \), \( S=9 \), and
\( P_{tot}=30\; kPa \) \cite{chap5:npentanol_thermal}.

The estimated values of the dimensionless parameters \( a \) and
\( b \) corresponding to those experimental conditions are \( a=5\, 10^{-8} \)
and \( b=8\, 10^{-12} \). As one can observe from Figs. \ref{pentanol a}
and \ref{pentanol b}, these small values of \( a \) and \( b \)
imply that neither diffusion not thermal diffusion have a significant
influence on the results of these particular experiments. This is
mainly due to the high values of both the equilibrium and the total
pressures under which these experiments are performed.

The situation changes drastically when either the equilibrium pressure
of the substance or the total pressure is low. To illustrate that
fact, we have repeated the previous calculation using the model compound
whose properties are listed in Tab. \ref{propiedades npentanol}.
Figures \ref{pentanol a} and \ref{pentanol b} illustrate the influence
of diffusion and thermal diffusion on the nucleation rate for this
model compound at \( T=293.15K \), \( S=12.5 \) \cite{chap5:becker,chap5:seinfeld}.
When nucleation takes place at a normal pressure, e.g. \( P_{tot}=1\; atm \),
the resulting value for the parameter controlling diffusion effects
is \( a=10^{-4}\phi  \), indicating again that diffusion is not especially
significant. However, this model compound is peculiar in having a
small saturation pressure. This indicates that if the total pressure
is low enough, diffusion effects can be important. In fact, in experiments
performed in thermal diffusion chambers, for reasons of stability
the chamber must be operated at reduced pressure to avoid convection
\cite{chap5:katz I,chap5:estabilidad}. Using a realistic bound of
\( P_{tot}<100p_{eq} \), one obtains that \( a>3 \) for \( P_{tot}<10^{-2}\; Torr \),
thus implying a reduction of the nucleation rate by three orders of
magnitude, as indicated in Fig. \ref{pentanol a}. This confirms that
at reduced pressures diffusion may have a significant influence on
the nucleation rate of substances. 

The properties of some pollutants commonly found at the atmosphere
are similar to those of the model compound \cite{chap5:becker}. Therefore,
the results of this section provide some insight into the possible
effect that diffusion may have in in low pressure upper atmospheric
processes.

\subsubsection{Condensation Experiments in Laminar Flow Diffusion Cloud Chambers}

The previous analysis remains valid for experiments performed in laminar
flow diffusion cloud chambers \cite{chap5:hameri laminar 1}-\cite{chap5:pentanol laminar}.
In these experiments, the imposition of a steady laminar flow is responsible
for the appearance of a region in which the temperature abruptly drops
and nucleation occurs. Once again, the zone in which nucleation takes
place is narrow, and temperature gradients are also present. (Representative
temperature, supersaturation, and nucleation rate profiles occurring
in these devices are shown for instance in Fig. 6 of Ref. \cite{chap5:laminar analysis}).
Therefore, we can use the results of the previous section to assess
the significance of diffusion and thermal diffusion in these experiments.

To be more precise, consider a typical experiment using n-pentanol
as a nucleating substance. From Ref. \cite{chap5:pentanol laminar},
representative conditions for these experiments are \( T=260K \),
\( S=10 \), and \( P_{tot}=100\; kPa \). Introducing the properties
of n-pentanol tabulated in Tab. (\ref{propiedades npentanol}) into
Eqs. (\ref{a}) and (\ref{b}), yields the values \( a=4\; 10^{-9}\phi  \),
and \( b=8\; 10^{-12}\psi  \) for the dimensionless parameters. These
very small values of \( a \) and \( b \) indicate that, under those
conditions, these experiments in laminar flow diffusion cloud chambers
are not significantly affected by the loss of subcritical clusters
through diffusion or thermal diffusion.

\subsubsection{Influence of diffusion and thermal diffusion in Polymer Crystallization}

A simplified model of polymer crystallization can be proposed in terms
of the coordinates of the center of mass of a spherical polymer crystallite
and the number of monomers it contains. Under those assumptions, the
analysis developed in the previous section remains valid for the description
of the polymer crystallization process.

Although the theoretical description of condensation and crystallization
can be similar, the crystallization process has some features that
makes its description more difficult\cite{chap1:kn:Kelton}. For the
case of crystallization, and especially for polymer crystallization,
the identification of the transport coefficients is not so straightforward.
Moreover, the attachment rate \( k^{+}(n) \) and especially the surface
tension are not really known. For this reason it is very difficult
to arrive at a detailed quantitative estimate of the diffusion and
thermal diffusion effects in the case of polymer crystallization.
Nevertheless, we present here a qualitative discussion of the probable
influence of such effects. 

In the case of crystallization, the rate \( k^{+}(n) \) at which
units attach to the growing crystal embryo, is roughly proportional
to the rate of diffusion of the units. This proportionality implies
that any change in the rate of diffusion in the system alters the
crystallization rate accordingly, since the mechanism of diffusion
is also responsible for the growth of the crystals. The importance
of this becomes evident for the case of crystallization under nonisothermal
conditions. When a temperature gradient is present in the system,
the diffusivity of monomers is changed by the appearance of thermal
diffusion, as indicated by Eq. (\ref{flujo espacial}). Moreover,
the total diffusivity is no longer isotropic but is different in the
direction determined by the temperature gradient. Since any reduction
or increase of the diffusivity leads to the corresponding reduction
/increase of the nucleation rate, this implies that the presence of
temperature gradients may significantly alter the nucleation rate.
We can estimate the potential magnitude of this effect. 

Thermal diffusion is proportional to the temperature gradient and
to the thermal diffusion coefficient \( D_{th} \), as indicated by
Eq. (\ref{flujo espacial}). \( D_{th} \) is in general smaller than
the normal diffusion coefficient in gases, but may become extremely
significant in polymers. A measure of the importance of thermal diffusion
follows from the ratio of thermal and diffusion coefficients, or Soret
coefficient

\begin{equation}
\label{Soret}
S=\frac{D_{th}}{D_{0}Tx_{1}(1-x_{1})}.
\end{equation}
Typical values of \( S \) for gaseous and liquid mixtures range from
\( 10^{-3} \) to \( 10^{-5} \) \( K^{-1} \) in order of magnitude,
the smallness of which indicates that mass diffusion is dominant.
For polymer solutions, however, the situation is completely different.
The Soret coefficient increases with the molecular weight to the extent
of becoming significant at high values of the molecular weight, as
reported in Ref. \cite{chap5:sengers}. Moreover, the small thermal
conductivity of the melted polymer allows the release of latent heat
involved in the crystallization process to set up very big temperature
gradients, a result that follows from Eq. (\ref{ecuacion del calor}).
Both factors imply that the magnitude of the thermal diffusion in
polymers under non-isothermal conditions is significant, and therefore
can influence the crystallization process importantly.

\section{Conclusions}

In this paper we have outlined some of the principles of mesoscopic
nonequilibrium thermodynamics (MNET) and have used this discipline
to establish the basis for a complete description of the kinetics
of nucleation in which the metastable phase acts as a heat bath in
which the embryos of the new phase are embedded. We have explicitly
shown that the presence of the heat bath may play a significant role
in the nucleation process.

To illustrate this, we discussed, in particular, the influence of
inhomogeneities in the medium on the process of nucleation by analyzing
in detail the process of nucleation in the presence of a temperature
gradient. We derived a kinetic equation, of the Fokker-Planck type,
that governed the evolution of the cluster size distribution, as it
was coupled to the evolution of the heat bath. This equation was then
used to evaluate the influence of the media in two particular situations
of practical interest, namely, condensation nucleation experiments
in diffusion chambers, and polymer crystallization.

The results of our analysis suggest that, when experiments are performed
at normal conditions, condensation in thermal and laminar flow diffusion
cloud chambers is not significantly affected by diffusion and thermal
diffusion effects. However, in rarefied media, as in upper atmosphere
or for substances with low equilibrium vapor pressures, those mechanisms
can become quite significant. 

Another situation in which the nucleation process can be drastically
influenced by the presence of a temperature gradient occurs in the
case of polymer crystallization. Polymers usually have low thermal
conductivities, which favors the development of very large temperature
gradients, and high values of the Soret coefficient. Since both factors
control the importance of thermal diffusion, this is a clear sign
that these effects may become crucial in polymer crystallization.
Finally, it is worth mentioning that diffusion and thermal diffusion
effects have proved to be important also in chemical vapor deposition,
as analyzed in Ref. \cite{chap5:seinfeld}. 

The framework proposed in this paper can be extended to more general
situations, such as the one in which the system is subjected to a
shear flow. This situation will be treated in the accompanying paper,
Paper II of this series. The method we have introduced thus goes beyond
the classical formalism and can account systematically for the real
conditions under which nucleation takes place.

\begin{acknowledgments}
The authors would like to acknowledge Prof. H. Reiss for his comments
and his careful revision of the manuscript, and I. Santamar\'{\i}a-Holek
for valuable discussions. This work has been partially supported by
the National Science Foundation under NSF grant No. CHE-0076384, and
by \foreignlanguage{english}{DGICYT of the Spanish Government under
grant PB2002-01267.}
\selectlanguage{english}

\newpage
\end{acknowledgments}

\section*{List of References}

\newpage

\section*{List of Tables}

\begin{itemize}
\item \textbf{Table I:} \foreignlanguage{american}{Thermophysical properties
of n-pentanol and the model compound (ethane sulfinic acid) using
helium as a carrier gas. The properties are: \( M \), the molar weight;
\( p_{eq} \), the saturation pressure; \( \rho _{l} \), the liquid
density; \( \sigma  \), the surface tension; \( M_{c} \), the molar
weight of the carrier gas (helium); \( \alpha _{T} \), is the thermal
diffusion factor; and \( x_{1} \) is the mole fraction. In the table,
\( T \) represents the absolute temperature in degrees Kelvin and
\( Z=1-T/588.15 \). Data taken from Refs. \cite{chap5:becker} and
\cite{chap5:npentanol_thermal}.\label{propiedades npentanol}}
\end{itemize}
\newpage

\section*{List of Figures}

\begin{itemize}
\item \textbf{Figure 1:} \foreignlanguage{american}{Schematic supersaturation
profile in a diffusion cloud chamber.}
\item \textbf{Figure 2:} \foreignlanguage{american}{Ratio between the rate
of nucleation with diffusion, \( J(n^{*}) \), and without diffusion,
\( J_{0} \), for n-pentanol at \( T=260K \), \( S=9 \) (heavy line)
and for the model compound (dashed line) as a function of the dimensionless
parameter \( a \) (and \( b=0 \)).}
\item \textbf{Figure 3:} \foreignlanguage{american}{Ratio between the rate
of nucleation in the presence of thermal diffusion, \( J(n^{*}) \),
and in its absence, \( J_{0} \), for n-pentanol at \( T=260K \),
\( S=9 \) (heavy line) and for the model compound (dashed line) as
a function of the dimensionless parameter \( b \) (for \( a=0 \)).}
\end{itemize}
\newpage

\section*{Table I}

\vspace{0.5cm}\lyxline{\normalsize}\vspace{-1\parskip}
\selectlanguage{american}
{\raggedright \begin{tabular}{p{13.2cm}}
\selectlanguage{american}
n-pentanol
\selectlanguage{english}\\
\end{tabular}\par}

\scriptsize

\begin{LyXParagraphIndent}{3cm}
\lyxline{\normalsize}\vspace{-1\parskip}
{\centering \begin{tabular}{l}
\selectlanguage{american}
\hspace{1cm}{\small \( M=88.15\; g/mol \) }
\selectlanguage{english}\\
\selectlanguage{american}
{\small \hspace{1cm}\( p_{eq}=133.322\, \exp (90.079043-9788.384/T-9.9\log T)\; Pa \) }
\selectlanguage{english}\\
\selectlanguage{american}
{\small \hspace{1cm} \( \rho _{l}=0.270+1.930229\, Z^{1/3}-8.414762\, Z^{2/3}+19.226001\, Z-18.559303\, Z^{4/3} \)}
\selectlanguage{english}\\
\selectlanguage{american}
{\small \hspace{1cm}\( \; \; \; \; +6.555718\, Z^{5/3}\: g/cm^{3} \)}
\selectlanguage{english}\\
\selectlanguage{american}
{\small \hspace{1cm}\( \sigma =26.85469-0.07889\, (T-273.15)\; dyn/cm \)}
\selectlanguage{english}\\
\selectlanguage{american}
{\small \hspace{1cm}\( M_{c}=4.0026\; g/mol \) }
\selectlanguage{english}\\
\selectlanguage{american}
{\small \hspace{1cm}\( 1/\alpha _{T}=(-0.7272-T/(16.36-0.2882\, T))(x_{1}+0.12281)+0.089303 \)}
\selectlanguage{english}\\
\end{tabular}\par}

\end{LyXParagraphIndent}

\normalsize

\vspace{0.5cm}\lyxline{\normalsize}\vspace{-1\parskip}
{\raggedright \begin{tabular}{l}
\selectlanguage{american}
Model compound
\selectlanguage{english}\\
\end{tabular}\par}

\scriptsize

\begin{LyXParagraphIndent}{3cm}
\lyxline{\normalsize}\vspace{-1\parskip}
{\noindent \raggedright \begin{tabular}{l}
\selectlanguage{american}
\hspace{1cm}{\small \( M=94.13\; g/mol \) }
\selectlanguage{english}\\
\selectlanguage{american}
\hspace{1cm}{\small \( \rho _{l}=0.868\: g/cm^{3}\; (v_{1}=1.8\, 10^{-3}\; cm^{3}) \)}
\selectlanguage{english}\\
\selectlanguage{american}
{\small \hspace{1cm}\( \sigma =30.0\; dyn/cm \)}
\selectlanguage{english}\\
\selectlanguage{american}
{\small \hspace{1cm}\( p_{eq}=10^{-5}\; Torr \)}
\selectlanguage{english}\\
\end{tabular}\par}

\end{LyXParagraphIndent}

\vspace{2mm}\hrule\vspace{1mm}\hrule\normalsize
\selectlanguage{english}

\newpage

\newpage

\begin{figure}
\selectlanguage{american}
{\centering \resizebox*{0.9\columnwidth}{!}{\includegraphics{fig2.eps}} \par}

\caption{\mbox{}\label{pentanol a}}
\selectlanguage{english}
\end{figure}

\newpage
\begin{figure}
\selectlanguage{american}
{\centering \resizebox*{0.9\columnwidth}{!}{\includegraphics{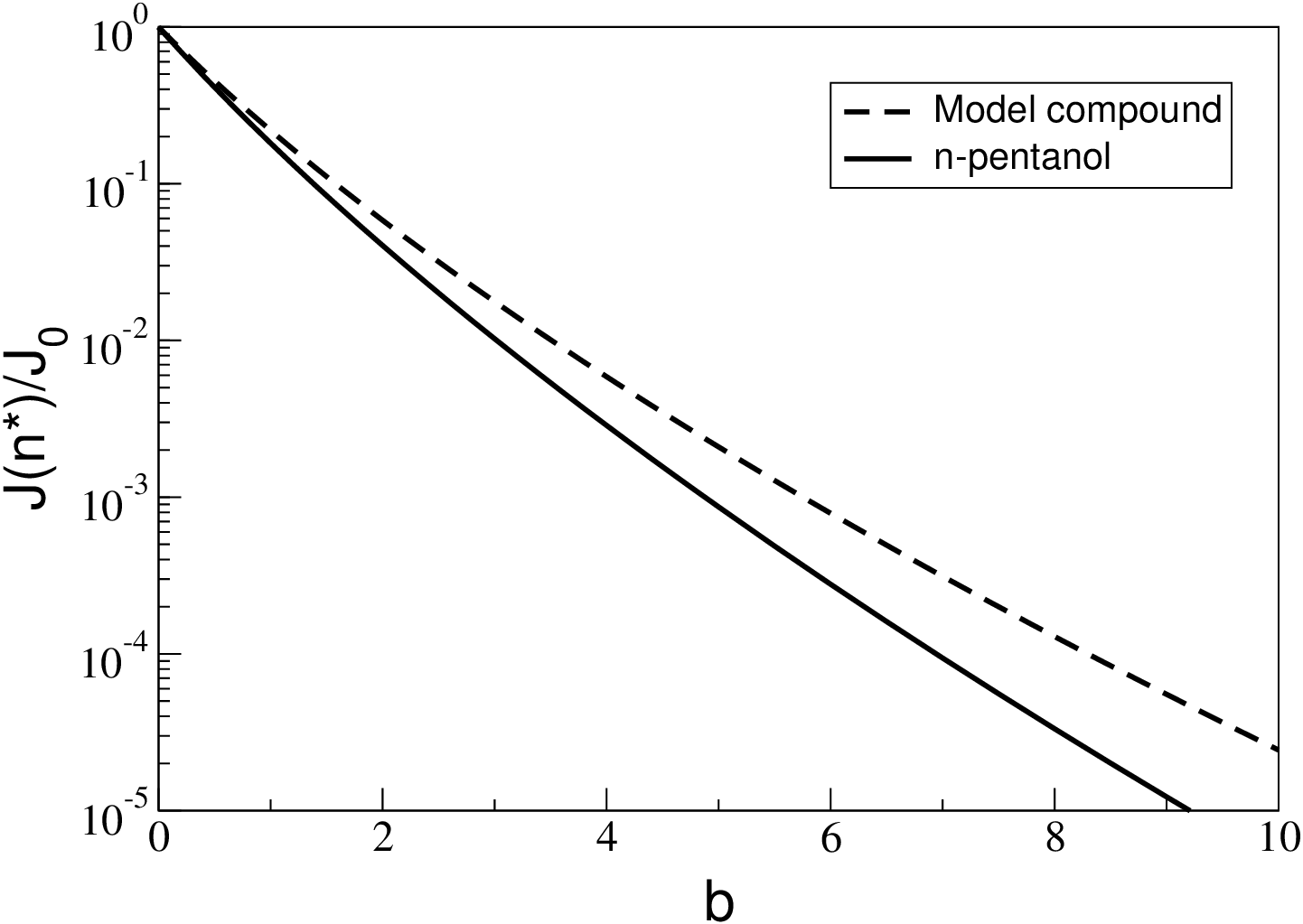}} \par}

\caption{\mbox{}\label{pentanol b}}
\selectlanguage{english}
\end{figure}

\end{document}